# On the density scaling of liquid dynamics

D. Fragiadakis and C. M. Roland

*Naval Research Laboratory, Chemistry Division, Code 6120, Washington, DC 20375-5342, USA*

Superpositioning of relaxation data as a function of the product variable $TV^\gamma$, where $T$ is temperature, $V$ the specific volume, and $\gamma$ a material constant, is an experimental fact demonstrated for approximately 100 liquids and polymers. Such scaling behavior would result from the intermolecular potential having the form of an inverse power law (IPL), suggesting that an IPL is a good approximation for certain relaxation properties over the relevant range of intermolecular distances. However, the derivation of the scaling property of an IPL liquid is based on reduced quantities, for example, the reduced relaxation time equal to $T^{1/2}V^{-1/3}$ times the actual relaxation time. The difference between scaling using reduced rather than unreduced units is negligible in the supercooled regime; however, at higher temperature the difference can be substantial, accounting for the purported breakdown of the scaling and giving rise to different values of the scaling exponent. Only the $\gamma$ obtained using reduced quantities can be sensibly related to the intermolecular potential.

Extensive measurements over the last few years of relaxation properties of organic materials at elevated pressures have led to many insights into the dynamics of glass-forming liquids and polymers [1,2]. An interesting aspect of these studies is the observation that over a wide range of thermodynamic conditions relaxation times conform to the scaling law

$$\tau = f(TV^\gamma) \quad (1)$$

where $f$ is a function, $V$ the specific volume, and $\gamma$ a material constant [3,4,5,6]. Eq (1) has been extended to other dynamical quantities, such as the viscosity $\eta$ [7,8] and diffusion coefficient $D$ [9], and also applied to molecular dynamics simulations (mds) [10,11,12,13,14,15] of particles interacting via Lennard-Jones (LJ) type potentials

$$U(r) = \varepsilon\left[\left(\frac{\sigma}{r}\right)^n - \left(\frac{\sigma}{r}\right)^6\right] \quad (2)$$

where $\sigma$ is the particle diameter, $\varepsilon/4$ the well depth, and $n$ a material constant. Of particular interest, mds results have demonstrated a connection between the $\gamma$ in eq. (1) and the steepness of the intermolecular repulsive potential (related to the $n$ in the repulsive term in eq. (2)) [12,15]. This is intriguing because the idea underlying density scaling is that the dynamic properties of viscous liquids are governed primarily by the repulsive component of the intermolecular potential, with changes in the long-range attractive part exerting a negligible effect. To the extent this is the case, for certain properties eq. (2) can be approximated by an inverse power law (IPL)

$$U(r) = \varepsilon\left(\frac{\sigma}{r}\right)^{-\tilde{n}} \quad (3)$$

Since eqs (2) and (3) are functions of distance but not orientation, such potentials can only be approximate for non-spherical molecules. However, the latter can be described using multiple, discrete LJ-site interactions [16,17,18], and such models exhibit correlation of fluctuations in potential energy and the virial [14]. This is consistent with an effective IPL, since the potential energy and the virial are perfectly correlated only for an IPL. This applicability of the IPL approximation to the supercooled dynamics, however, does not imply any general validity. For example, the melting behavior of non-spherical molecules exhibits strong deviations from density scaling [19].

The excess transport and thermodynamic properties ("excess" refers to the non-ideal part, dependent on particle positions but not momenta) of an IPL liquid are a function of $TV^{\tilde{n}/3}$ [20,21]. This follows from the fact that an IPL and its derivatives depend only on the combined variable $\varepsilon\sigma^{-\tilde{n}}$, so that properties depend only on a single temperature-density variable. The implication is that conformance of real materials to density scaling (eq. (1)) results from their intermolecular potential being approximately an IPL, at least for consideration of certain "local" dynamic properties. Dyre and coworkers [14,15] coined the term "isomorphic states" to refer to state points in which the partition function constructed from a given potential depends to a good approximation (or exactly for an IPL) on the single variable $TV^\gamma$.

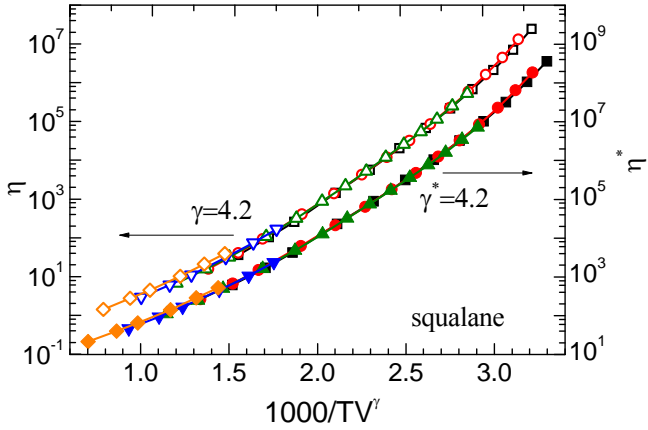

FIG. 1. Density scaling of viscosity (open symbols) and reduced viscosity (filled symbols) of squalane. The data extend over many decades of viscosity, therefore both quantities scale with identical scaling exponents. In this and all other figures $TV^\gamma$ is in units of K (ml/g)$^\gamma$ and viscosities are mPa s; reduced units are dimensionless.

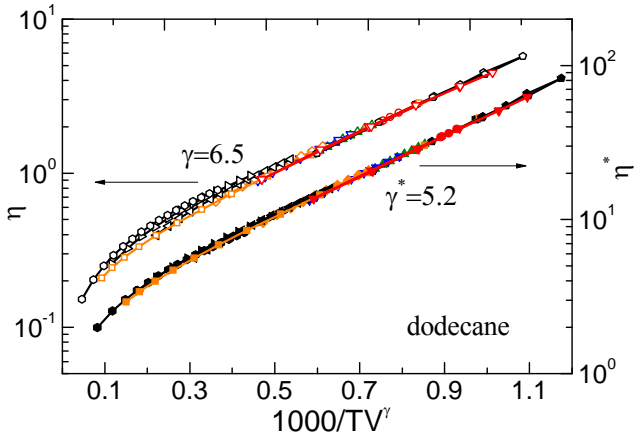

FIG. 2. Density scaling of viscosity (open symbols) and reduced viscosity (filled symbols) of dodecane. The reduced viscosity shows better scaling and a lower scaling exponent.

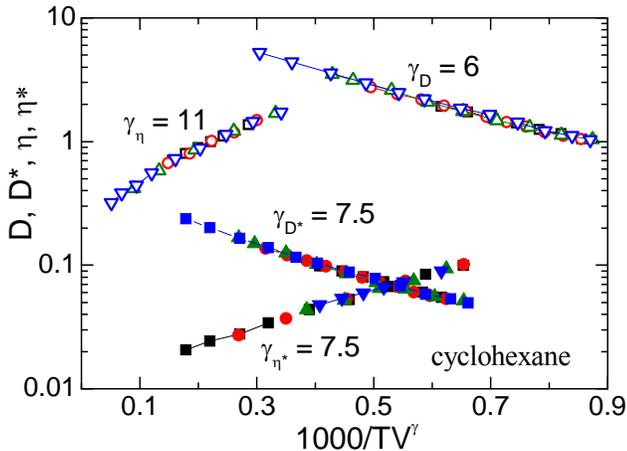

FIG. 3. Density scaling of the viscosity and diffusion coefficient of cyclohexane using unreduced (open symbols, $D$ in cm$^2$/sec) and reduced (filled symbols) units. For the latter the exponent is independent of the dynamic variable.

Although for a strict IPL $\gamma = \tilde{n}/3$, results for LJ liquids indicate that the scaling exponent is larger than the repulsive exponent ($\gamma > n/3$) [12], the steeper repulsive slope ($\tilde{n} > n$) due to the presence of the attractive term in eq. (2) [22]. Nevertheless, such results suggest that density scaling of experimental data has the potential to yield direct information about the forces between molecules in real materials. However, there is a disconnect between scaling of dynamic data from mds and from experiments. For the latter, scaling is applied to $\tau$, $\eta$, $D$ as measured, with the scaling property being an observable fact. However, the scaling property of effective IPL potentials applies to reduced quantities, using units of time $t_0 = v^{1/3}(kT/m)^{-1/2}$, length $l_0 = v^{1/3}$ and energy $E_0 = kT$, where $m$ and $v$ are molecular mass and volume [20,21]. Dimensionless reduced quantities are thus defined as

$$\tau^* = v^{-1/3}(kT/m)^{1/2}\tau$$
$$\eta^* = v^{2/3}(mkT)^{-1/2}\eta \quad (4)$$
$$D^* = v^{-1/3}(kT/m)^{-1/2}D$$

Accordingly, for an IPL and mds data the form of eq. (1) is

$$\tau^* = f(TV^{\gamma^*}) \quad (5)$$

with similar expressions for $\eta^*$ and $D^*$. Since over the range of experimental measurements on supercooled liquids, $T$ varies by as much as 100% and $v$ by ca. 10 %, it is not a priori obvious that differences between reduced and non-reduced units are negligible. Conformance of experimental data to eq. (5) and any difference between $\gamma^*$ and $\gamma$ are important issues if the scaling exponent is to be interpreted as the slope of an effective intermolecular potential.

Density scaling is carried out by plotting a dynamic property variable versus $TV^\gamma$, with $\gamma$ adjusted to collapse the data onto a master curve. If the variation in $\tau$, $\eta$, or $D$ is much greater than that of volume and temperature, then the shape of this master curve will be essentially identical whether actual or reduced quantities are employed. Since the hallmark of the supercooled state is the enormous change in molecular mobilities for small changes in thermodynamic conditions [2], scaling of $\tau$ or $\tau^*$ is practically equivalent. This is illustrated for squalane in Figure 1, which is representative of the literature results for dielectric relaxation times of supercooled liquids and polymers.

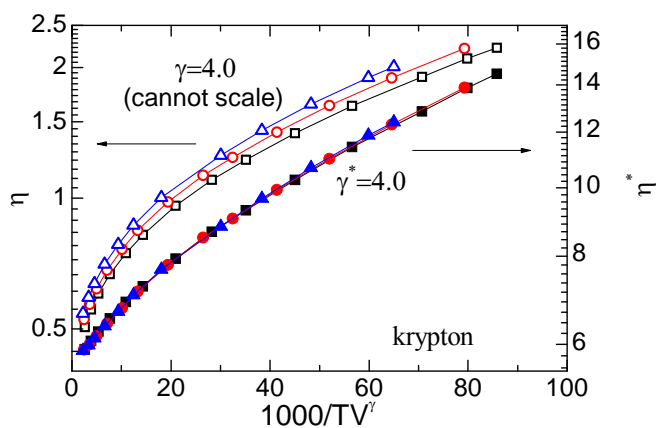

FIG. 4. Density scaling of viscosity (open symbols) and reduced viscosity (filled symbols) of krypton. The reduced viscosities scale well with an exponent $\gamma^*=4.0$; however, the scaling is poor for any value of $\gamma$.

Many liquids cannot be supercooled and relaxation data (typically viscosity or self-diffusion coefficient measured using NMR) are available over only a narrow range above the melting point. In such circumstances the use of reduced units makes a substantial difference. Figure 2 shows high-pressure viscosity data for a typical case, dodecane. Pensado et al. [8] reported a scaling exponent of $\gamma=6.5$ for the viscosity, with good superpositioning for smaller values of $TV^\gamma$, but the scaling deteriorated for lower $\eta$. Adjusting the value of $\gamma$ improves scaling for lower $\eta$, but then the higher viscosity data diverge. Alternatively, the scaled viscosities are plotted Figure 2, and the superposition is very good over the entire range with $\gamma^*=5.2$. Note that if the dynamic data encompassed a narrower range, as is often the case, both reduced and non-reduced viscosities would ostensibly scale equally well, albeit with different values of the scaling exponent. For example, for toluene $\gamma=7.8$ [7], whereas we find $\gamma^*=5.4$; for n-hexane $\gamma=13$ [8], while we find $\gamma^*=7$. Also, the scaling exponent $\gamma$ is independent of the experimental variable (relaxation time, transport coefficient, etc.) only when reduced units are used. This is illustrated in Figure 3 for the viscosity and diffusion constant of cyclohexane (using data from ref. [23]).

We can also apply the density scaling relation to very low viscosity liquids, such as krypton (shown in Figure 4), argon, nitrogen, oxygen, and methane (using viscosity data from ref. [24]). For these substances, non-reduced viscosities do not superpose versus $TV^\gamma$ for any value of $\gamma$. However, the reduced viscosities conform to the scaling relation, with physically reasonable values of $\gamma^*=4.0$ for krypton, argon, nitrogen and methane, and 5.0 for oxygen.

The above results affirm the connection between $\gamma$ and the steepness of the intermolecular potential for $r$ relevant to the local dynamics. Density scaling is not merely an empirical relation useful for organizing experimental data, but has a physical basis in the IPL approximation for the intermolecular potential. If the data encompass a range sufficient to yield different values for $\gamma$ and $\gamma^*$, the latter, based on reduced dynamic quantities, is the physically relevant exponent. Examples for which $\gamma^*$ differs from published $\gamma$ include n-alkanes [8], octane and toluene [7], 1-alkylamines [25] and 2-alkylamines [26], carbon dioxide and toluene [9]. Moreover, some examples of an ostensible breakdown of the scaling are a consequence of using non-reduced variables.

## ACKNOWLEDGMENTS

This work was supported by the Office of Naval Research. DF acknowledges a National Research Council post-doctoral fellowship.

## REFERENCES

[1] C.M. Roland, S. Hensel-Bielowka, M. Paluch, R. Casalini, *Rep. Prog. Phys.* 68, 1405 (2005).
[2] C.M. Roland, *Macromolecules* 43, 7875 (2010).
[3] A. Tolle, *Rep. Prog. Phys.* 64, 1473 (2001).
[4] R. Casalini, C.M. Roland, *Phys. Rev. E* 72, 031503 (2005).
[5] C. Alba-Simionesco, A. Cailliaux, A. Alegria, G. Tarjus, *Europhys. Lett.* 68, 58 (2004).
[6] C. Dreyfus, A. Le Grand, J. Gapinski, W. Steffen, A. Patkowski, *Eur. J. Phys.* 42, 309 (2004).
[7] C.M. Roland, S. Bair and R. Casalini, *J. Chem. Phys.* 125, 124508 (2006).
[8] A.S. Pensado, A.A.H. Pádua, M.J.P. Comuñas and J Fernández, *J. Phys. Chem. B* 112, 5563 (2008).
[9] K.R. Harris, *J. Chem. Phys.* 131, 054503 (2009).
[10] J. Budzien, J.D. McCoy, and D.B. Adolf, *J. Chem. Phys.* 121, 10291 (2004).
[11] G. Tsolou, V.A. Harmandaris, and V.G. Mavrantzas, J. Chem. Phys. 124, 084906 (2006).
[12] D. Coslovich, C.M. Roland, *J. Phys. Chem. B* 112, 1329 (2008).
[13] D. Coslovich, C.M. Roland, *J. Chem. Phys.* 131, 151103 (2009).
[14] N.P. Bailey, U.R. Pedersen, N. Gnan, T.B. Schrøder, and J.C. Dyre, *J. Chem. Phys.* 129, 184507 (2008); ibid. 184508 (2008).
[15] T.B. Schrøder, U.R. Pedersen, N.P. Bailey, S. Toxvaerd, and J.C. Dyre, *Phys. Rev. E* 80, 041502 (2009).
[16] W.B. Streett and K.E. Gubbins, *Ann. Rev. Phys. Chem.* 28, 373 (1977).
[17] V.N. Kabadi and W.A. Steele, *J. Phys. Chem.* 89, 1467 (1985).
[18] F.J. Vesely, *J. Chem. Phys.* 125, 214106 (2006).
[19] D. Fragiadakis and C.M. Roland, *Phys. Rev. B, submitted* (2010).
[20] W.G. Hoover, M. Ross, *Contemp. Phys.* 12, 339 (1971).

[21] Y. Hiwatari, H. Matsuda, T. Ogawa, N. Ogita, and A. Ueda, *Prog. Theor. Phys.* 52, 1105 (1974).
[22] D. Ben-Amotz and G. Stell, *J. Chem. Phys.* 119, 10777 (2003).
[23] J. Jonas, D.H. Hasha, S.G. Huang, *J. Chem. Phys.* 84, 109 (1980).
[24] NIST Chemistry WebBook, *NIST Standard Reference Database*, Number 69, ed. by P. J. Linstrom and W. G. Mallard, NIST, Gaithersburg, 2005, URL http://webbook.nist.gov.
[25] M. Yoshimura, C. Boned, A. Baylaucq, G. Galliéro and H. Ushiki, *J. Chem. Thermodynamics* 41, 291 (2009).
[26] M. Yoshimura, C. Boned, G. Galliéro, J.-P. Bazile, A. Baylaucq and H. Ushiki, *Chem. Phys.* 369, 126 (2010).